\documentstyle[11pt,paspconf]{article}
\def\beq{\begin{equation}}
\def\eeq{\end{equation}}
\def\ref{\reference}
\def\simge{\mathrel{%
   \rlap{\raise 0.511ex \hbox{$>$}}{\lower 0.511ex \hbox{$\sim$}}}}
\def\simle{\mathrel{
   \rlap{\raise 0.511ex \hbox{$<$}}{\lower 0.511ex \hbox{$\sim$}}}}

\def\edcomment#1{\iffalse\marginpar{\raggedright\sl#1\/}\else\relax\fi}
\reversemarginpar

\begin{document}
\title{Advection-Dominated Accetion Flows}
\author{Insu Yi$^{1,2,3,4}$}
\affil{$^1$Institute for Advanced Study, Princeton, NJ 08540, USA; yi@ias.edu}
\affil{$^2$Center for High Energy Astrophysics and Isotope Studies, 
Research Institute for Basic Sciences}
\affil{$^3$Physics Department, Ewha University, Seoul, Korea; 
yi@astro.ewha.ac.kr}
\affil{$^4$Korea Institute for Advanced Study, Seoul, Korea}

\begin{abstract}
We review basic properties of advection-dominated accretion flows (ADAFs)
and their applications to astrophysical systems ranging from Galactic 
binary systems to galactic nuclei. A new classification scheme for
low-luminosity, X-ray bright galactic nuclei is highlighted. Some outstanding 
unresolved issues are discussed.
\end{abstract}

\section{Introduction}
It has recently been recognized that rotating accretion flows with 
low radiative efficiency are applicable to a wide range of astrophysical
systems including Galactic X-ray transients and active galactic nuclei.
In the recently discussed models for such sources, the radiative efficiency
becomes low because the gravitational binding energy dissipated during infall 
of accreted matter is not efficiently radiated away due to low gas densities 
but stored within the flows and radially advected inward. Viscous torque 
transports angular momentum while generating heat through dissipation. 
Radiative efficiency is basically determined by the fraction of the
viscously dissipated energy which goes into radiation (Narayan \& Yi 1995b). 
In the optically thin limit, 
which occurs when the density of the accretion flow falls below a
certain limit, the radiative efficiency becomes small due to the long time
scales for electron-ion energy exchange process and relevant cooling processes.
The first discussions of such flows are found in Ichimaru (1977),
Rees et al. (1982), and references therein.

The accretion flows around compact objects could be classified into several
types (e.g. Narayan et al. 1998b, Frank et al. 1992).
(i) Geometrically thin disks radiate away a large fraction 
of the viscously dissipated energy and due to high density, the optical
depth for outgoing radiation is large. The disk temperature is relatively low
and hence the internal pressure support is small, which results in the
small geometrical thickness (e.g. Frank et al. 1992 for review). In these
flows, the cooling rate $Q^-$ is balanced by the heating rate $Q^+$,
$Q^+=Q^-$ while the electron temperature $T_e$ equals the ion temperature
$T_i$. (ii) Two-temperature, Shapiro-Lightman-Eardley (1976) type accretion
flows also maintain the energy balance as $Q^+=Q^-$ while 
$T_i\sim T_{vir}\gg T_e$ where $T_{vir}$ is the usual virial temperature. These
flows are thermally unstable (Piran 1978, cf. Rees et al. 1982). 
(iii) Slim disk or optically thick advection-dominated accretion flows 
(ADAFs) occur when accretion
rates typically exceed the Eddington rate and the optical depth is large.
In this limit, photons diffuse out on a time scale longer than the radial
inflow time scale (Abramowicz et al. 1988 and references therein). These
flows have $Q^+>Q^-$ and $T_i=T_e$. 
(iv) Optically thin ADAFs (Ichimaru 1977, Rees et al. 1982, 
Narayan \& Yi 1994, 1995b, Abramowicz et al. 1995) are
relevant for substantially sub-Eddington accretion rates. The optical depth in
these flows are typically less than unity and the radiative cooling rate is
much smaller than the heating rate, i.e. $Q^+>Q^-$, 
while the ion temperature
$T_i\sim T_{vir}$ is much higher than the electron temperature $T_e$. These
flows are most interesting in systems which have relatively low luminosities
and high emission temperatures (Narayan \& Yi 1995b).

In this review, we focus on the optically thin ADAFs and discuss their 
applications to various astrophysical systems.

\section{Basics of ADAFs: A Simple Version}

\subsection{Basic Equations}

Following Narayan \& Yi (1994, 1995ab), we adopt the following notations
for a steady, axisymmetric, rotating accretion flow; $R=$ cylindrical radius 
from the central star, $\rho=$ gas density, $H=$ thickness or vertical scale 
height of the flow, $v=$ radial velocity, $\Omega=$ angular velocity,
$\Omega_K=(GM/R^3)^{1/2}=$ Keplerian angular velocity, $c_s=$ isothermal 
sound speed, $\nu=$ kinematic viscosity coefficient, $T=$ is temperature of 
the gas, and $s=$ specific entropy of the gas. Then, the basic conservation
equations for mass, radial momentum, angular momentum, and energy become
respectively,
\beq
\rho R H v=constant,
\eeq
\beq
v{dv\over dR}-(\Omega^2-\Omega_K^2)R=-{1\over \rho}{d\over dR}(\rho c_s^2),
\eeq
\beq
\rho R H v{d(\Omega R^2)\over dR}={1\over dR}\left(\nu\rho R^3 H
{d\Omega\over dR}\right),
\eeq
\beq
\rho v T {ds\over dR}=q^+-q^-\equiv fq^+
\eeq
where
$q^+=\nu\rho R^2(d\Omega/dR)^2=$ viscous dissipation rate per unit volume,
$q^-=$ radiative cooling rate per unit volume, and
$\rho v T(ds/dR)=q^{adv}=$ radial advection rate per unit volume. 
Therefore, the energy equation
becomes simply $q^{adv}=q^+-q^-=fq^+$
which defines the advection fraction $f$. When advection cooling per unit
volume dominates,
$q^-\ll q^+$, $f\approx 1$. 
Even when $f$ differs substantially from unity, most of dynamical 
calculations assume a constant f rather than solving the full energy equation 
(or assumes a simple cooling such as bremsstrahlung).
Viscosity coefficient is specified by the $\alpha$ prescription
(e.g. Frank et al. 1992) $\nu=\alpha c_s H=\alpha c_s^2/\Omega_K$
where $\alpha$ is a constant often assumed to be in the range $0.01-1$,

The accretion flows are classified into three types according to the relative
importance of the terms in the energy equation (Narayan et al. 1998).
(i) $q^+\approx q^-\gg q^{adv}$: energy balance is maintained between
viscous heating and radiative cooling, which corresponds to high-efficiency
flows such as geometrically thin disks. 
(ii) $q^+\approx q^{adv}\gg q^-$: radiative loss is negligible and 
luminosity is very low. (iii) $q^+\ll q^-\approx -q^{adv}$: viscous heating is 
negligible and thermal energy of the flow is converted to radiation as in the
cooling flows. Infall of matter is driven by pressure loss as gas cools.

\subsection{Self-Similar Solution}

The dynamical equations derived admit a self-similar solution for $f=$constant
as shown in Narayan \& Yi (1994) and Spruit et al. (1987);
\beq
v=-\left[{(5+2\epsilon^{\prime})\over 3\alpha^2}g(\alpha,\epsilon^{\prime})
\right]\alpha v_K,
\eeq
\beq
\Omega=\left[{2\epsilon^{\prime}(5+2\epsilon^{\prime})\over 9\alpha^2}
g(\alpha,\epsilon^{\prime})\right]^{1/2} {v_K\over R},
\eeq
\beq
c_s^2=\left[{2(5+2\epsilon^{\prime})\over 9\alpha^2}
g(\alpha,\epsilon^{\prime})\right]v_K^2,
\eeq
where
$v_K=R\Omega_K=(GM/R)^{1/2}$, $\epsilon^{\prime}=\epsilon/f$, 
$\epsilon=(5/3-\gamma)/(\gamma-1)$, and
$g(\alpha,\epsilon^{\prime})= \left[1+{18\alpha^2\over 
(5+2\epsilon^{\prime})^2}\right]^{1/2}-1$. The specific heat ratio
$\gamma=4/3-5/3$.

For $\alpha=0.01-0.3$, $\alpha^2\ll 1$ and $f\approx 1$ gives
\beq
v\approx -\left(9\gamma-9\over 9\gamma-5\right)\alpha v_K,
\eeq
\beq
\Omega\approx \left[2(15-9\gamma)\over 3(9\gamma-5)\right]^{1/2}\Omega_K\le
\Omega_{K},
\eeq
\beq
c_s^2\approx {6\gamma-6\over 9\gamma-5}v_K^2
\eeq
The self-similar solution reveals the basic properties of the ADAFs. (i)
The radial accretion time scale is much shorter than that of the thin disk.
(ii) Sub-Keplerian rotation occurs due to large internal pressure support.
(iii) Vertical scale height $H\sim c_s/\Omega_K\sim R$. Moreover, the
positive Bernoulli parameter indicates that ADAFs are prone to outflows
although there have not been any self-consistent inflow/outflow solutions
(Narayan \& Yi 1995a).
The vertically integrated equations and the self-similar solutions do not
introduce any serious errors in flow dynamics since the height integration 
is a good approximation (Narayan \& Yi 1995a).

\subsection{Cooling and Heating Mechanisms}

In ADAFs, due to low radiative cooling efficiency of electrons and
rather ineffective ion-electron coupling, which is taken to be the Coulomb
coupling, ions are nearly virialized (Narayan \& Yi 1995b)
\beq
T_i\sim 2\times 10^{12}\beta r^{-1} K.
\eeq
Electrons' energy balance is maintained as (Mahadevan \& Quataert 1997)
\beq
\rho T_e v {ds\over dR}=\rho v {d\epsilon\over dR}-kT_v{dn\over dR}=q^{ie}
+\delta q^+-q^-
\eeq
where $kT_ev(dn/dR)=q^{compress}$ is the compressive heating and 
$\delta q^+$ is 
the direct viscous heating on electrons with $\delta\sim m_e/m_p\sim 10^{-3}$ 
(Nakamura et al. 1996). $q^{ie}>q^{compress}$ for ${\dot m}\simge 0.1\alpha^2$
and $\rho v (d\epsilon/dR)\approx q^{ie}-q^-$ describes the electron energy
balance. $q^{ie}<q^{compress}$ occurs when ${\dot m}\simle 10^{-4}\alpha^2$ and
the energy balance for electrons becomes $\rho v (d\epsilon/dR)\approx 
q^{compress}$, which is appropriate only for very low luminosity systems.
$\delta q^+< q^{compress}$ is realized only for $\delta\simle 10^{-2}$ which
implies that the direct viscous heating is uninteresting in most practical
cases (e.g. Mahadevan \& Quataert 1997).

For convenience, we introduce some physical scalings;
mass $m\equiv M/M_{\odot}$, radius $r=R/R_s$ ($R_s=2GM/c^2=2.95\times 
10^5 m~cm$), accretion rate ${\dot m}={\dot M}/{\dot M}_{Edd}$ 
(${\dot M}_{Edd}=L_{Edd}/0.1c^2=1.39\times 10^{18} m~ g/s$ where $L_{Edd}$
is the Eddington luminosity). The equipartition 
magnetic field $B^2/8\pi=(1-\beta)\rho c_s^2$ with $\beta=0.5$
and $f=1$ where $\beta$ is the ratio of magnetic to total pressure.
The self-similar solution gives the physical quantities in terms of the
dimensionless variables defined here.
\beq
v\approx -1\times 10^{10} \alpha r^{-1/2}~cm/s,
\eeq
\beq
\Omega\approx 3\times 10^4 m^{-1} r^{-3/2}~s^{-1},
\eeq
\beq
c_s^2\approx 1\times 10^{20} r^{-1}~cm^2/s^2,
\eeq
\beq
n_e\approx 6\times 10^{19} \alpha^{-1} m^{-1} {\dot m} r^{-3/2}~cm^{-3},
\eeq
\beq
B\approx 8\times 10^8 \alpha^{-1/2} m^{-1/2} r^{-5/4} {\dot m}^{1/2}~G,
\eeq
\beq
\tau_{es}\approx 24\alpha^{-1} r^{-1/2} {\dot m},
\eeq
\beq
q^+\approx 5\times 10^{21} m^{-2} r^{-4} {\dot m}~erg/s/cm^3,
\eeq
where $n_e$ is the electron number density and $\tau_{es}$ is the electron
scattering depth.

Various (electron) cooling processes give rise to distinct spectral 
components (Narayan \& Yi 1995b).
The integrated cooling rate $Q\sim\int dV q$ where $q$ is the cooling rate 
per unit volume and $\int dV$ denotes integration over the entire accretion
flow. The total electron cooling rate is
\beq
Q^-_e=Q_{sync}+Q_{Compt}+Q_{brem}
\eeq
where $Q_{sync}$ is the synchrotron cooling rate which gives rise to
spectral emission components in radio, IR, or optical/UV depending on the
mass $m$ and accretion rate ${\dot m}$. $Q_{Compt}$ is the Compton cooling
rate which is mainly responsible for optical/UV/soft X-ray emission.
$Q_{brem}$ is the bremsstrahlung cooling contributing to X-ray and
soft gamma-ray emission. If ADAFs in the inner regions around accreting
compact objects are surrounded by the optically thick disks, optical/UV 
emission from cool disks is expected. 
Energetic protons in ADAFs may result in high energy
gamma-rays (Mahadevan et al. 1997). 
Similar radiation processes in zero angular momentum spherical 
accretion have been extensively discussed (e.g. Melia 1992 and references).

\subsection{Critical Quantities}

ADAFs exist when accretion rates fall below a certain critical rate
${\dot M}_{crit}={\dot m}_{crit}{\dot M}_{Edd}$. Such a critical rate
arises because there exists a maximum accretion rate above which heating 
could be balanced by radiative cooling without any necessity of advective
cooling  (Rees et al. 1982, Narayan \& Yi 1995b, Abramowicz et al. 1995,
Narayan et al. 1998). 
In the case of the single temperature case, i.e. $T_e=T_i\propto r^{-1}$, 
assuming an optically thin, bremsstrahlung cooling 
(good for $r>10^3$), 
we get $q^+\propto m^{-2} r^{-4} {\dot m}\propto {\dot m}$ and
$q^-=q^-_{brem}\propto \rho^2 T_e^{1/2}\sim \rho^2 T\propto \alpha^{-2}
m^{-2} r^{-7/2} {\dot m}^2\propto {\dot m}^2$ or from $q^+\sim q^-$ we get
the critical accretion rate ${\dot m}_{crit}\propto \alpha^2 r^{-1/2}$.
In the case of the single temperature with $T_e=T_i\ne T_{vir}$, assuming 
synchrotron and Compton cooling, the critical accretion rate becomes
${\dot m}_{crit}\sim 10^{-4}\alpha^2$. These well motivated
critical rates are too low to be of practical interest (Esin et al. 1997). Esin
et al. (1997) found that the critical rate deduced from observed spectral
transition in soft X-ray transients is much higher than the above rates.

In the two-temperature ADAFs, the bottleneck in energy transfer from ions
to electrons define another critical rate which is good for $r\simle 10^3$
(e.g. Narayan et al. 1998).
That is, using $q^+\propto {\dot m}$ and $q^-\propto q^{ie}\propto {\dot m}^2$
and equating the two rates, $q^+=q^{ie}$, gives a critical accretion rate
${\dot m}_{crit}\approx 1\times 10^3 (1-f) \epsilon^{\prime} \alpha^2 
\beta^{-1/2} r^{-1/2}$. Alternatively, $t_{ie}=t_{acc}\approx R/v$ gives 
${\dot m}_{crit}\approx 0.3\alpha^2$.
In sum, ${\dot m}_{crit}\sim \alpha^2$ for $r\simle 10^3$ and 
${\dot m}_{crit}\propto \alpha^2 r^{-1/2}$ for $r\simge 10^3$, which depicts
the actual radial dependence of the critical accretion rate.

It is interesting to point out that there exists a critical $\alpha$ 
(Chen et al. 1995). For $\alpha<\alpha_{crit}\sim r$, ${\dot m}_{crit}$ 
exists while for $\alpha>\alpha_{crit}\sim r$, ${\dot m}_{crit}$ doesn't exist.

\subsection{Some Recent Works on Heating Ions}

Bisnovatyi-Kogan \& Lovelace (1997) recently claimed that large electric fields
parallel to magnetic fields can accelerate electrons and hence bypassing the
bottleneck in energy transfer from ions to electrons, which could rule out
ADAFs as a possible accretion flow type. However, such a possibility is 
realized  only when a substantial resistivity on microscopic scale exists. 
This requires a small magnetic Reynolds number. Their proposal to use the 
macroscopic turbulent resistivity is not applicable on microscopic scales 
as pointed out by Blackman (1998) and Quataert (1998).
Blackman (1998) argued that the Fermi acceleration by large scale magnetic 
fluctuations associated with MHD turbulence may lead to preferential ion 
heating and hence two-temperature plasma. This heating is not applicable to
non-compressive Alfvenic turbulence which is most likely to be more important
than the compressive mode.
For weak magnetic fields substantially weaker than equipartition fields, 
Alfvenic component of MHD turbulence is dissipated on scales of proton Larmor 
radii (Gruzinov 1998, Quataert 1998). This mechanism favors ion heating and
two-temperature plasma. For strong fields (i.e. near equipartition), 
the Alfvenic turbulence cascades to scales much smaller than proton Larmor 
radii and can directly heat electrons, which could cast doubt on ADAFs with 
equipartition strength magnetic fields. That is, there is a possibility that 
equipartition plasma doesn't allow two-temperature plasma. This issue may 
ultimately be resolved by observed spectra.

\subsection{ADAF Luminosity}
In ADAFs, the observed radiative luminosity
$L_{ADAF}=\int L(E) dE\sim \int q^{ie} dV$ or in terms of the ADAF
radiative efficiency $\eta_{ADAF}$, $L_{ADAF}=\eta_{ADAF} {\dot M} c^2$
where
$\eta_{ADAF}=\eta_{eff}\times 0.2 {\dot m} \alpha^{-2}\propto {\dot m}\propto
{\dot M}/M$ (Narayan \& Yi 1995b). In contrast, the thin disk
luminosity $L_{thin\quad disk}\sim \eta_{eff}{\dot M} c^2$
with $\eta_{eff}\sim 0.1$ (e.g. Frank et al. 1992).

\section{Some Physical Issues}

\subsection{Global Solutions}

The self-similar solution considered so far applies to regions far from
the inner and outer boundaries where physical scales of the system demand the
dynamical equations to deviate from self-similarity.
A comprehensive summary of the Newtonian case is found in Kato et al. (1998).
The task of finding the global solution is to find an eigenvalue $j$ 
(specific angular momentum accreted by the central object) with proper 
boundary conditions. The proper boundary conditions are
(i) outer thin disk matching inner ADAF,
(ii) sonic point for the transonic ADAF, and
(iii) vanishing torque at the inner boundary.
Solving the radial momentum equation, angular momentum equation, and
energy equation, along with the implicit continuity equation for $v, \Omega, 
c_s, \rho$ and eigenvalue $j$, global solutions are found.
The self-similar solution is a good approximation for a wide range of radii
between the inner and outer boundaries. The pseudo-Newtonian potential case 
has been solved by Matsumoto et al. (1985), Narayan et al. (1997), 
Chen et al. (1997). The major findings are as follows.
(i) For $\alpha\simle 0.01$, inefficient angular momentum transport results in
slow radial accretion and a wide radial zone of super-Keplerian rotation.
In the super-Keplerian rotation region, a thick torus-like structure 
with funnel 
around the rotation axis forms and the pressure profile shows a maximum, 
which is reminiscent of the ion torus model (e.g. Rees et al. 1982).
(ii) For $\alpha\simge 0.01$, efficient angular momentum transport and 
rapid accretion $v\propto \alpha v_K$ occur. There exist no pressure maximum
and the accretion flows are quasi-spherical.

In the relativistic case with a spinning black hole (Abramowicz et al. 1996, 
Peitz \& Appl 1997, Gammie \& Popham 1998, Popham \& Gammie 1998), the
Newtonian and pseudo-Newtonian results are largely confirmed for
$r\simge 10$. For $r\simle 10$, however, significant differences and spin 
effects are seen. Detailed calculations of emission spectra have been
carried out by Jaroszynski \& Kurpiewski (1997).

It has been an issue whether shocks form in transonic accretion flows.
In the steady calculations, shocks are not seen and in the time-dependent
calculation of Igumenshchev et al. (1996), no shocks have been seen.
Manmoto et al. (1996)'s time-dependent calculation shows shock-like 
steepening in density waves, which could be responsible for ADAF variabilities.

\subsection{Stability}

The geometrically thin disks could be thermally and viscously unstable
under certain circumstances (e.g. Frank et al. 1992). ADAFs are primarily
stable agains thermal and viscous perturbations: 
(i) ADAFs are stable against long wavelength perturbations (Narayan \& Yi 1995b,
Abramowicz et al. 1995). (ii) ADAFs are marginally stable against short 
wavelength perturbations in the single temperature case
(Kato et al. 1996, 1997, Wu 1997).
(iii) ADAFs are stable against short wavelength perturbations both 
thermally and viscously in the two temperature case (Wu \& Li 1996).
The stability of ADAFs have led to an argument that the thermally
unstable Shapiro-Lightman-Eardley disk (Piran 1978) is a spatially
transitional accretion flow linking the thermally unstable outer thin disk 
to the stable inner ADAF.
In the time-dependent calculations, small wavelength perturbations in
the single temperature ADAFs have been observed to  grow while moving
inward. The growth rate is however not rapid enough to affect the
steady global structure (Manmoto et al. 1996).

\section{X-ray Transients}
\subsection{Black Hole Systems}

Detailed spectral fitting has been tried for black hole systems such as 
A0620-00, V404 Cyg (Narayan et al. 1996), Nova Mus 1991, Cyg X-1, 
GRO J0422+32, GRO J1719-24 (Esin et al. 1997), and 1E1740.7-2942
(Vilhu et al. 1997). The main physical parameters used in the spectral
fitting are $M$, ${\dot m}$, $\alpha$, $\beta$, and $r_{tr}$, where the
last one is the radius of accretion flow transition from outer thin disk
to inner ADAF. The spectral fitting assumes that (i) the outer thin disk is
joined to the inner ADAF  and (ii) the outer thin disk is unstable against
disk instability. The instability causes the heating/cooling waves to
propagate inward, resulting in delays between different emission components 
(Lasota et al. 1996, Hameury et al. 1997). The transition radius $r_{tr}$ 
is crucial in determining spectra (Lasota et al. 1996, Narayan et al.
1996,1998, Esin et al. 1998).

\subsection{Thin Disk - ADAF Transition}
ADAFs exist when $q^-<q^+$ which in principle determines $r_{tr}$.
In the single temperature, bremsstrahlung dominated case,
\beq
q^+_{vis}\propto m^{-2}r^{-4}{\dot m}\propto r^{-4}
\eeq
\beq
q^-_{brem}\propto \rho^2 T^{1/2}\propto\alpha^{-2}m^{-2}r^{-7/2}{\dot m}^2
\propto r^{-7/2}
\eeq
and $q^+=q^-$ gives (Honma 1996)
\beq
r_{tr}\sim 3\times 10^2 (\alpha^4/{\dot m}^2).
\eeq
In more realistic cases, details of transition are unknown. For instance,
Honma (1996) considers the radial turbulent heat diffusion near the interface
between the thin disk and the ADAF. 
Spectral fitting gives a different result on $r_{tr}$
(Esin et al.  1997). The issue of the accretion flow transition still
remains unresolved.

\subsection{Neutron Star Systems}

Neutron star transients have not been well studied using the ADAF models mainly
due to the lack of sufficiently well-developed physical understanding of
accretion flows near the neutron star surface. The radiation feedback
from the soft radiation emitted by the stellar surface makes the spectral
calculations considerably more complicated (Narayan \& Yi 1995b,
Yi et al. 1996). Moreover, it is unclear whether the spectral transition
similar to that of black hole systems exists in neutron star systems.

\subsection{Accretion-Powered X-ray Pulsars}

Spectral transition in neutron star systems could be potentially very difficult
to detect if thermalization near the neutron star surface occurs rapidly
(e.g. Yi et al. 1996). Accretion-powered X-ray pulsars could however provide
an observable signature of the temporal accretion flow transition. 
In accretion-powered pulsars such as 4U 1626-67, GX 1+4, and OAO 1657-415,
abrupt torque reversals have been observed, which are extremely difficult to 
explain within the conventional models involving disk-magnetospheric
interaction (Chakrabarty et al. 1997). The difficulties arise mainly due to
(i) short reversal time scales, (ii) nearly identical spin-up and spin-down 
rates, (iii) small X-ray luminosity changes, (iv) significant spectral 
transition reported in 4U 1626-67 (Vaughan \& Kitamoto 1997), and
(v) torque-flux correlation observed in GX 1+4 (Chakrabarty et al. 1997).

The observed phenomenon is well accounted for by the accretion disk transition 
triggered by a gradual, small amplitude modulation of mass accretion rate. The 
sudden reversals occur at a rate $\sim 10^{16-17} g/s$ when the accretion flow 
makes a transition from (to) a primarily Keplerian flow to (from) a 
substantially sub-Keplerian, radially advective flow (Yi et al. 1997, 
Yi \& Wheeler 1998). The proposed transition model naturally
shows that (i) the transition time scale is likely to be shorter than days,
(ii) the required accretion rate change is at the level of a few $\times 10$ 
percent, and (iii) the abrupt reversal is a signature of a pulsar system near 
spin equilibrium with small mass accretion rate modulations near the critical 
accretion rate on the time scale of years. Other possible explanations 
for the spectral transition and the torque-flux correlation have been 
suggested (Nelson et al. 1997, van Kerkwijk et al. 1998) with varying
difficulties in explaining the above observational facts. The accretion flow 
transition is similar to those in black hole transients and cataclysmic 
variables, which strongly suggests a common physical origin (Yi et al. 1997).
The transition time scale is $t_{thermal}\sim (\alpha\Omega_K)^{-1}$ or
$t_{vis}\sim R/\alpha c_s\sim (R/H)t_{thermal}$, where the latter time
scale becomes $\sim 10^3s$ for $\alpha\sim 0.3$, $R\sim 10^9 cm$, 
${\dot M}\sim 10^{16} g/s$. This time scale is short enough to induce 
transition on a time scale of a day or less.

If the accretion flow's thermal pressure is a substantial fraction $\xi$ of the
dynamical pressure, i.e. $c_s^2\sim \xi R^2 \Omega_K^2$, the thickness of
the flow $H\sim \xi^{1/2}R$, the radial accretion speed 
$v_R\sim -\alpha\xi R\Omega_K$, and the angular rotational frequency 
$\Omega\sim (1-5\xi/2-\alpha^2\xi^2/2)^{1/2}\Omega_K=A\Omega_K$
where $\xi\rightarrow 0$ is the Keplerian limit and $A< 1$ for ADAFs and
$A=1$ for thin Keplerian disks. When $A<1$ occurs, $R_c^{\prime}=A^{2/3}R_c$,
where $R_c$ is the Keplerian corotation radius and $R_c^{\prime}$ is the 
sub-Keplerian corotation radius.
The accretion flow is truncated at a radius $R_o^{\prime}$ and
using $N_0^{\prime}=A{\dot M}(GM_*R_o^{\prime})^{1/2}$ the torque on the star
becomes
\beq
{N^{\prime}\over N_0^{\prime}}={7\over 6}
{1-(8/7)(R_o^{\prime}/R_c^{\prime})^{3/2}\over
1-(R_o^{\prime}/R_c^{\prime})^{3/2}}
\eeq
which pushes the pulsar's spin toward an equilibrium spin
$P_{eq}^{\prime}=P_{eq}/A>P_{eq}$ where ${\prime}$ denote quantities after
transition from thin disk to ADAF. In this picture, torque reversal is
expected if
\beq
B_*\sim 5\times 10^{11} L_{x,36}^{1/2}P_{*,10}^{1/2}G
\eeq
where $L_{x,36}=L_x/10^{36} erg/s$ is the X-ray luminosity and 
$P_{*,10}=P_*/10s$ is the pulsar spin period. 
Observed quasi-periodic oscillation (QPO) periods tightly constrain the
proposed model. Yi \& Grindlay (1998) discuss some possible implications 
on spin-up of LMXBs to MSPs when the accretion flows are ADAFs in these
systems.

\subsection{Energetic Protons: Lithium Production}

Energetic ions present in ADAFs are capable of nuclear spallation. Lithium
production in ADAFs in binary systems and ion tori  in galactic nuclei
has been discussed by Ramadurai \& Rees (1985), Jin (1990), 
Yi \& Narayan (1997). Using the self-similar solution, the relevant
physical quantities are as follows: number density of protons
$n_H\sim 6\times 10^{20} m^{-1} {\dot m} r^{-3/2} cm^{-3}$, number density
of $\alpha$ particles 
$n_{\alpha}\sim 5\times 10^{19} m^{-1} {\dot m} r^{-3/2} cm^{-3}$,
energy per nucleon $E\sim 300r^{-1} MeV$, and the radial accretion speed
$v_R\sim 2\times 10^9 r^{-1/2} cm/s$. The production of ${^7}Li$ dominates
and the production cross-section $\sigma_+(E)\sim 100(E/100 MeV)^{-2} mbarn$
for $E\ge 8.5MeV$. Continuous production of $^7Li$ within the accretion flow
leads to increase in Lithium abundance according to
\beq
{\Delta n_{Li}\over n_H}={1\over 2}\sigma_+(E)v_r{n_{\alpha}^2\over n_H}
\Delta t_{flow}
\eeq
where $\Delta t_{flow}=\Delta R/v_R$. The enrichment results in the terminal
abundance $n_{Li}/n_H=\int d (n_{Li}/n_H)\approx 0.1{\dot m}$ or the
total Lithium mass production rate
${\dot M}_{Li}=2\times 10^{-8} m {\dot m}^2 M_{\odot}/yr$.

It is expected that $^7Li$ enrichment occurs around NSs and BHs containing 
ADAFs, which has been seen in recent precision spectroscopic measurements of
$^7Li$ in V404 Cyg, A0620-00, GS 2000+25, Nova Mus 1991, Cen X-4  
(Martin et al. 1994 and references therein). 
In contrast, WD systems show no such effect,
which indicates that  only at $r\sim 1$ relativistic energies are reached
while $r\sim 10^3$ is the inner most radius in the WD systems, which gives 
$E\ll 1MeV$.

\subsection{Thermalization of Particles}

Although we have adopted the thermal temperatures for ions and electrons,
it is not proven that particle energy distributions are adequately
approximated by thermal distributions. Recent investigations
(e.g. Mahadevan \& Quataert 1997, Quataert 1998, Blackman 1998) have suggested
some limited information on this unresolved issue.

In most of the ADAF models, protons and ions are energized by viscous heating
which is mostly unspecified. Alfvenic turbulence (which does not result in 
strongly non-thermal distributions) and Fermi acceleration 
(leading to power-law tails) have been considered. Coulomb collisions and 
synchrotron absorption do not lead to rapid thermalization of protons. So far
it appears that the acceleration mechanism itself determines the proton
energy distributions. Thermalization of electrons could occur more easily
(Ghisellini \& Svensson 1991). Coulomb collisions can thermalize
electrons for ${\dot m}\simge 10^{-2}\alpha^2$. Synchrotron self-absorption
leads to thermalization when ${\dot m}\simge 10^{-5}\alpha^2 r$. The electron
energy distributions directly affect radio emission spectra.

Recently Mahadevan (1998) pointed out that low frequency $\nu\simle 10^9Hz$ 
radio spectrum of Sgr A$^*$ could be contributed by electrons and positrons
produced by charged pions from proton-proton collisions and that
neutral pion production could account for tentative detection of gamma-rays
in the direction of Sgr $A^*$. In both cases, particle distributions need to
be strongly nonthermal and the results depend very sensitively on high energy 
tails.

\section{Galactic Nuclei}

ADAFs may exist in galactic nuclei including the Galactic center source
Sgr $A^*$. For our discussions, we define $m_7=m/10^7$, 
${\dot m}_{-3}={\dot m}/10^{-3}$, and $R_s=2GM/c^2=3\times 10^{12} m_7 cm$.

When the accretion flow is a thin disk, the luminosity $L=\eta {\dot M}c^2$
with the efficiency $\eta=\eta_{eff}\sim 0.1$ (e.g. Frank et al. 1992). 
Although the total luminosity
is high, the emission temperature or the disk temperature
$T_{disk}\sim 6\times 10^6 m_7^{-1/5}{\dot m}_{-3}^{3/10} r^{-3/4} K$ is
too low to account for X-ray emission. The disk luminosity  
$L_{disk}\sim 1\times 10^{42} m_7 {\dot m}_{-3} erg/s$ is expected to occur in
optical/UV/soft X-ray. X-ray emission and radio emission are usually accounted
for by optically thin corona and radio jets. For the latter, the estimated
radio power is $L_{jet}\sim 1\times 10^{42} {\bar a}^2 \eta_{jet} m_7 
{\dot m}_{-3} erg/s$ where ${\bar a}$ is the black hole spin parameter and
$\eta_{jet}$ is the jet radiative efficiency.

In ADAFs, relevant physical quantities scaled for galactic nuclei are
equipartition magnetic field $B\sim 1\times 10^4 m_7^{-1/2}{\dot m}_{-3}^{1/2}
r^{-5/4}G$, electron scattering depth 
$\tau_{es}\sim 5\times 10^{-2} {\dot m}_{-3}$, the ion temperature
$T_i\sim 2\times 10^{12} r^{-1}~K$, and the electron temperature
$T_e\sim 5\times 10^9~K$. ADAFs are expected to exist when the mass accretion
rate falls below ${\dot m}_{crit}={\dot M}_{crit}/{\dot M}_{Edd}\approx 0.3
\alpha^2 \sim 10^{-3}-10^{-2}$.
Radio emission is easily explained by the synchrotron emission
with the characteristic synchrotron emission frequency (Yi \& Boughn 1998ab)
\beq
\nu_{sync}\sim 1\times 10^{12} m_7^{-1/2} {\dot m}_{-3}^{1/2} r^{-5/4} 
T_{e9}^2~Hz
\eeq
where $T_{e9}=T_e/10^9K\sim 5$. The highest synchrotron radio emission
frequency is $\nu_{max}=\nu_{sync}(r\sim 1)\sim 3\times 10^{13} m_7^{-1/2} 
{\dot m}_{-3}^{1/2}~Hz$ which comes from the inner most region of ADAF near
the black hole horizon. The radio luminosity 
\beq
L_R\sim \nu L_{\nu}^{sync}\sim 2\times 10^{32} x_{M3}^{8/5} T_{e9}^{21/5}
m_7^{6/5} {\dot m}_{-3}^{4/5}\nu_{10}^{7/5}~erg/s
\eeq
where $x_{M3}=x_M/10^3$ is a dimensionless synchrotron self-absorption
parameter and $\nu_{10}=\nu/10^{10}~Hz$. ADAF radio emission
could be tested by the radio source size - frequency relation. ADAFs predict
such a relation 
\beq
\theta(\nu)
\sim 2 m_7^{3/5} {\dot m}_{-3}^{2/5}\nu_{10}^{-4/5} (D/10Mpc)^{-1}~{\mu}as
\eeq
where the angular size $\theta(\nu)$ becomes $\sim mas$ for distance scales
$D\sim 10 kpc$.

Optical/UV/X-ray in ADAFs arise from inverse Compton scattering of radio
synchrotron photons and hard X-rays are from bremsstrahlung and multiple
Compton scattering. For ${\dot m}\simle 10^{-3}$, X-ray emission is
dominated by the bremsstrahlung emission, $L_x\sim L_x^{brem}\propto m 
{\dot m}^2$ and the radio luminosity $L_R\propto m^{8/5}{\dot m}^{6/5}$, which
suggests $L_R\propto mL_x^{3/5}$. For ${\dot m}\simge 10^{-3}$, X-ray emission
gives the luminosity $L_x^{Compt}\propto {\dot m}^{7/5+N}$ with $N\ge 2$.
In this case, we expect $L_{R}\propto mL_x^{6/5(N+1)}$. 
Yi \& Boughn (1998ab) have derived and tested the radio/X-ray luminosity 
relation for ADAFs using $L_x=L_x(2-10 keV)$;
\beq
L_R\sim 10^{36} m_7 (\nu/15GHz)^{7/5}(L_x/10^{40} erg/s)^x~erg/s
\eeq
where $x\sim 1/5$ for ${\dot m}\simle 10^{-3}$ and $x\sim 1/10$ for
${\dot m}>10^{-3}$ or $L_{R,adv}/L_{x,adv}\propto m L_{x,adv}^{-1}$.

ADAFs are likely to drive jets/outflows (Narayan \& Yi 1995a). If jets are
powered by black hole's spin energy (e.g. Frank et al. 1992),
\beq
L_{R,jet}/L_{R,adv}\sim 4\times 10^5 {\bar a}^2 \eta_{jet} m_7^{-1/5}
{\dot m}_{-3}^{1/5}
\eeq
and $L_{R,jet}/L_{x,adv}\propto {\bar a}^2 m L_{x,adv}^{-1}$ are expected.
That is, $L_{R,jet}\gg L_{R,adv}$ for ${\bar a}\gg 2\times 10^{-3} 
\eta_{jet}^{-1/2}$. It is often argued that radio-loud nuclei have 
${\bar a}\simle 1$. If a galactic nucleus contains a thin disk and a jet
$L_{R,jet}/L_{x,disk}\sim {\bar a}^2\epsilon_{jet}/\eta_{eff}\sim O(1)$
is expected.

Characteristic ADAF emission spectra are determined primarily by
${\dot m}$ and weakly affected by the black hole mass $M$. 
Any combinations among $L_x$, $L_R$ and $M$ give useful information
on the nature of emission from galactic nuclei.

\subsection{Galactic Center Source Sgr A$^*$}

Galactic center radio source Sgr $A^*$ is a prime candidate for an
ADAF (Rees 1982, Narayan et al. 1995, 1998, Manmoto et al. 1997).
Sgr A$^*$, which is at the dynamical center of the Galaxy, appears to
contain a massive black hole with mass $\sim (2.5\pm 0.4)\times 10^6M_{\odot}$.
Wind accretion from nearby IRS 16 wind is expected to provide a mass
accretion rate ${\dot M}\simge$ a few $\times 10^{-6}M_{\odot}/yr$. With the
conventional $\sim 10\%$ efficiency, such a high accretion rate would
correspond to a luminosity of $0.1{\dot M}c^2\simge 10^{40} erg/s$ which is
some 3 to 4 orders of magnitude larger than the observed
radio to gamma-ray luminosity of $\simle 10^{37} erg/s$. Moreover, a
standard thin disk would give peak emission in near infrared but
Menten et al. (1997)'s 2.2 micron upper limit rules out this possibility.
These facts strongly suggest that an ADAF is present in Sgr $A^*$.

Spectral fitting based on the observed $M$ and the estimated ${\dot m}$ 
adequately accounts for the observed emission seen from radio to hard X-ray.
Bower \& Backer (1998) measured the intrinsic source size $\simle 0.48~mas$ 
at 7mm, which at 8.5kpc gives the linear size of 4.1AU and the lower limit 
on the brightness temperature $4.9\times 10^9K$. Both the size and temperature
measurements are consistent with the ADAF models. Inverted radio spectrum with 
sharp cutoff along with no jet like elongations near Sgr A$^*$ are also
consistent with the ADAF predictions.
X-ray constraints are rather uncertain. ROSAT 0.8-2.5 keV luminosity is
$\sim 1.6\times 10^{34} erg/s$ (Predehl \& Trumper 1994) and ASCA 2-10 keV 
luminosity is $\le 4.8\times 10^{35} erg/s$ (Koyama et al. 1996). Both 
constraints are easily satisfied by an ADAF.

\subsection{NGC 4258}

NGC 4258 almost certainly has a central black hole with mass 
$M=3.5\pm 0.1 \times 10^7M_{\odot}$, which is concentrated within 0.13pc 
(for a distance of 6.4Mpc) of the dynamical center.
This source has an observed 2-10 keV X-ray luminosity
of $L_x=4\times 10^{40} erg/s$ (Makishima et al. 1994). Optical luminosity 
$L_{opt}\simle 10^{42} erg/s$ (Wilkes et al. 1994) provides an additional
constraint. 22 GHz continuum emission (after subtraction of jet component) has
not been detected with a 3$\sigma$ upper limit of 220$\mu$Jy 
(Herrnstein et al. 1998) or a luminosity upper limit $L_R(22GHz)<2.4\times
10^{35} erg/s$. The non-detection of the core at 22 GHz could imply that
${\dot m}\sim 10^{-2}$ and $r_{tr}\sim 30$ if an ADAF exists in NGC 4258
(Gammie et al. 1998, cf. Lasota et al. 1996).
Such a constraint is highly suspect due to a possibility
of strong variabilities in radio emission from ADAFs (Blackman 1998, 
Ptak et al. 1998, Herrnstein et al. 1998).

\subsection{M60, M87, NGC1068, and M31}

Di Matteo \& Fabian (1997b) has attempted to fit M60 emission spectra with
an ADAF under the assumption that the accretion rate is close to the Bondi 
accretion rate with $M\sim 10^9 M_{\odot}$. Due to uncertainties and lack of
flux measurements, a definitive conclusion as to whether an ADAF exists needs
more data in X-ray or other wavebands.
Reynolds et al. (1996) claimed that an ADAF model spectrum for a blackhole
mass $M=3\times 10^9M_{\odot}$ and the accretion rate ${\dot m}\sim 10^{-3}$
accounts for the observed fluxes of M87. However, such a conclusion is highly
suspect because the extended radio emission component has not been properly
removed. ADAFs themselves do not produce extended radio emission.
NGC1068 is an obscured Seyfert with a very high obscuration-corrected
X-ray luminosity. X-rays are most likely to be dominated by scattering and
the X-ray luminosity may be too bright for ADAFs based on the estimated black 
hole mass $\sim$ a few$\times 10^7 M_{\odot}$ (Yi \& Boughn 1998ab).
Yi \& Boughn (1998b) also considered M31 which has a central black hole of
mass $M=3\times 10^7M_{\odot}$. The observed radio luminosity is too low 
for the observed X-ray luminosity if an ADAF is assumed around the black hole.
It is highly likely that the X-ray luminosity is dominated by binary sources 
in the nucleus while the radio emission may be due to a very weak ADAF.

\subsection{X-ray Bright Galactic Nuclei}

Yi \& Boughn (1998ab) and Franceschini et al. (1998) applied the ADAF
model to a small sample of X-ray bright galactic nuclei which have
black hole mass estimates. Since ADAFs are most relevant for low luminosity,
hard X-ray sources, faint, hard X-ray galactic nuclei are likely hosts of
ADAFs (Fabian \& Rees 1995, Di Matteo \& Fabian 1997a, Yi \& Boughn 1998ab).
Hard spectrum, faint X-ray sources could contribute significantly to
the diffuse X-ray background. 50\% of 2-10 keV XRB could be accounted for
by ADAF sources with the comoving density of $3\times 10^{-3} Mpc^{-3}$
for $L_x\sim 10^{41} erg/s$, which is comparable to the local density of 
the $L_*$ galaxy (Di Matteo \& Fabian 1997a, Yi \& Boughn 1998a).
However, unless $M\simge 10^9M_{\odot}$, $L_x\simge 10^{40} erg/s$ would 
be already too bright for ADAFs to account for the observed X-ray background.
This is because at high luminosities X-ray emission is dominated by the
Compton scattering which result in X-ray spectra much different from the
background spectrum similar to the bremsstrahlung-dominated X-ray spectra. 
That is, in relatively bright ADAFs, high ${\dot m}$'s correspond to the
Compton-dominated cooling regime. Although a significant clumping of ADAF 
gas would enhance bremsstrahlung over Compton (Di Matteo \& Fabian 1997),
such a possibility is difficult to realize.

So far, we have argued that ADAF models in galactic nuclei are testable
due to distinguishing characteristics of ADAFs. 
ADAFs with low radiative efficiency and high temperature are likely for 
massive black holes accreting at accretion rates $\simle 10^{-2}M_{Edd}$.
Hard X-ray emission and inverted spectrum radio emission from 
compact core are expected. There exists a characteristic radio/X-ray 
luminosity relation as shown by Yi \& Boughn (1998ab). For known black hole 
masses, existence of hot ADAFs can be tested by radio/X-ray observations. 
Black hole masses could be estimated based on radio/X-ray luminosities.
ADAF sources could however contain jets/outflows which can contribute to radio 
emission. 
Depending on the level of radio activity and existence of extended radio
emission features, galactic sources could be classified (Yi \& Bough 1998ab).
Such a classification can be quantified in a manner similar to that adopted 
for Galactic X-ray sources. In fact, there exist interesting spectral 
similarities between Galactic binary X-ray sources and galactic nuclei.

We define X-ray bright galactic nuclei (XBGN) as galactic nuclei with
X-ray luminosities in the range $10^{40}\simle L_x\simle 10^{42} erg/s$ which
is sub-luminous compared with the more powerful active galactic nuclei
(AGN) which generally have $L_x\simge 10^{43} erg/s$. Most of XBGN are
expected to overlap with emission line galaxies with $L_x\sim 10^{39}-10^{42} 
erg/s$. However, some of the low luminosity Seyferts with 
$L_x\simge 10^{42} erg/s$ cannot be ruled out. Based on our discussions
of ADAFs, for $L_x\sim 10^{41} erg/s$, 
\beq
L_R\sim 4\times 10^{36} (M_{BH}/3\times 10^7M_{\odot})~erg/s
\eeq
at 20 GHz with the characteristic inverted radio spectrum 
$I_{\nu}\propto \nu^{2/5}$. These sources at distances 
$\sim 10(M/3\times 10^7M_{\odot})^{1/2} Mpc$ should be detected as
$\sim 1 mJy$ point-like radio sources (Yi \& Boughn 1998a).
If X-ray and radio are indeed from ADAFs, the black hole masses can be 
estimated (Yi \& Boughn 1998b).

Yi and Boughn (1998ab) proposed the source classification based on the
known black hole masses and the ADAF radio/X-ray luminosity relation.
Adopting Sgr A$^*$, NGC 4258, NGC 1068, NGC 1316, NGC 4261, and NGC 4594
as fiducial sources, a statistically incomplete sample of XBGN are
classified into radio-loud XBGN and radio-quiet XBGN.
The former show that the observed radio
luminosity $L_{R,obs}\sim L_{R,jet}\gg L_{R,adv}$ where $L_{R,jet}$ and 
$L_{R,adv}$ are the expected radio jet luminosity and ADAF radio luminosity,
respectively. These sources are expected to show extended radio emission,
unlikely to have strongly inverted radio spectra, and may have compact ADAF 
radio emission from compact cores separate from the extended emission
components. The latter show that $L_{R,obs}\sim L_{R,adv}$ and that the
dominating emission components are compact cores with inverted spectra.

Kellermann et al. (1998) and Falcke (1998) show that jet-like radio 
emission features are common among AGN and emission line galaxies. Surprisingly,
even in radio-quiet sources, elongated radio emission features are sometimes 
seen on small scales, which could imply that some type of jet/outflow
activities are very common in galactic nuclei regardless of their large
scale radio activities. In order to resolve this issue, high resolution radio 
measurements for nearby ($<10Mpc$) sources are crucial.
Hard X-ray emission and inverted spectrum, compact radio emission are very 
likely to be found closely correlated.

\subsection{QSO Evolution}

Yi (1996) has suggested that transition of
accretion flows from thin disks to ADAFs at 
${\dot m}_{crit}\approx 0.3\alpha^2$ could account for the observed
sudden decline in the number of bright QSOs at redshift $z\sim 2$ and
downward (see also Fabian \& Rees 1995). 
Once ADAFs set in, the luminosity evolves according to
\beq
L=L_{ADAF}\approx 30{\dot m}^x L_{Edd}
\eeq
where $x\sim 2$ and ${\dot m}={\dot M}/{\dot M}_{Edd}\propto {\dot M}/M$.
the last expression implies that even when the mass accretion rate 
${\dot M}$ is kept constant, ${\dot m}$ decreases merely due to the growth of 
black hole mass.

For instance, in a flat universe with no cosmological constant, for 
the initial black hole mass $M=M_i$ at $z=z_i$ and ${\dot M}=constant$,
when ${\dot m}<{\dot m}_{crit}$ and ${\dot M}/H_o\gg M_i$
\beq
L(z)\propto (1+z)^{3(x-1)/2}\left[(1+z)^{-3/2}-(1+z_i)^{-3/2}\right]
\eeq
or
\beq
L(z)\propto (1+z)^{K(z)}
\eeq
with
\beq
K(z)={3(x-1)\over 2}\left[1+{1\over [(1+z_i)/(1+z)]^{3/2}-1}\right]
\eeq
which shows that the luminosity declines with a power-law similar to
that seen in observations (Yi 1996 and references therein).
The epoch at which ADAFs set in for ${\dot m}={\dot m}_i\sim 1$ is
(i.e. ${\dot m}={\dot m}_{crit}$ first occurs)
\beq
1+z_c=\left[\left(t_{Edd}\over t_o\right)\left({1\over {\dot m}_{crit}}-
{1\over \delta}\right)+(1+z_i)^{-3/2}\right]^{-2/3}
\eeq
where $\delta={\dot M} t_{Edd}/M_i$ and $t_{Edd}=({\dot M}_{Edd}/M)^{-1}=
4.5\times 10^7 (\eta_{eff}/0.1) yr$. The observed sudden decline of QSOs
at $z\sim 2$ is naturally accounted for.

\subsection{Outflows}

ADAFs are prone to outflows or jets (Narayan \& Yi 1994,1995a) although a
self-consistent inflow/outflow solution has not been found yet
(cf. Xu \& Chen 1997). It remains to be seen if a self-consistent 
inflow/outflow solution can account for compact and extended jet-like
emission components in XBGN.

\section{Some Unresolved Issues}

ADAFs may exist in sources spanning many decades of masses of compact
accreting sources. Some of the old outstanding issues, which are
mostly concerning with low-luminosities and hard X-ray emission, 
are plausibly resolved by various versions of ADAF models. 
There exist however a number of unsolved problems in the ADAF framework.

(i) The physics of accretion flow transition, temporal and spatial, remains
unclear. The spatial transition (i.e. from outer thin disk to inner
ADAF) is better understood than the temporal transition to a certain extent.
However, there has not been an adequate explanation for $r_{tr}$. It remains
unsolved why the disk flow makes a transition to ADAF with little, if any,
change of accretion rate.
(ii) Even luminous systems show energetic X-ray emission which is absent in
the thin disk models. It is often assumed that these sources have X-ray emitting
coronae for which little physics is known. It is crucial to link the ADAF
models to corona models with proper physical understanding.
(iii) ADAFs emission could be highly variable. The issue of steady vs.
non-steady ADAFs is directly related to the observed variabilities in ADAF
candidate sources which show occasional non-detections.
(iv) A number of plasma astrophysical issues remain to be solved. Is the
two-temperature flow physically allowed? Are particles rapidly thermalized?
What is the correct strength of magnetic fields responsible for synchrotron
emission? (v) Observationally, faint X-ray sources, which could be seen by
high resolution, high sensitivity experiments such as AXAF, should be
studied in great details. 
Ultimately, ADAF-related issues and the very question on the future of ADAFs' 
are likely to be answered by observations.

\end{document}